\documentclass[aps,prb,twocolumn,groupedaddress,showpacs]{revtex4}

\usepackage{graphicx}
\usepackage{dcolumn}
\usepackage{bm}

\usepackage{thumbpdf} 
\usepackage{hyperref}
\hypersetup{
    pdfstartview=FitH,
    pdfpagemode=UseThumbs,
    pdfauthor = {M. J. Mehl, A. Aguayo, L. L. Boyer and R. de Coss},
    pdftitle = {Absence of Metastable States in Strained Monatomic
    Cubic Crystals},
    pdfsubject = {Magic Strains in Crystals},
    pdfkeywords = {Elastic Constants, Magic Strains, Orthorhombic
    Strains, Density functional theory, Tight-binding},
    pdfcreator = {LaTeX with hyperref package},
    pdfproducer = {pdflatex}}

\newcolumntype{.}{D{.}{.}{-1}} 

\begin{document}

\title{Absence of Metastable States in Strained Monatomic Cubic
Crystals}

\author{Michael J. Mehl}
\email{mehl@dave.nrl.navy.mil}

\author{Aar\'{o}n Aguayo}
\altaffiliation{also at School of Computational Sciences, George Mason
University, Fairfax VA 22030}
\email{aguayo@dave.nrl.navy.mil}

\author{Larry L. Boyer}
\email{boyer@dave.nrl.navy.mil}
\affiliation{Center for Computational Materials Science,
Naval Research Laboratory, Washington, D.C. 20375-5345}

\author{Romeo de Coss}
\email{decoss@mda.cinvestav.mx}
\affiliation{Departamento de F{\'\i}sica Aplicada, Centro de
Investigaci\'on y de Estudios Avanzados, A.P. 73 Cordemex 97310
M\'erida, Yucat\'an, M\'exico}

\date{\today}

\begin{abstract}
A tetragonal (Bain path) distortion of a metal with an fcc (bcc)
ground state will initially cause an increase in energy, but at some
point along the Bain path the energy will again decrease until a
local minimum is reached.  Using a combination of parametrized
tight-binding and first-principles LAPW calculations we show that
this local minimum is unstable with respect to an elastic
distortion, except in the rare case that the minimum is at the bcc
(fcc) point on the Bain path.  This shows that body-centered
tetragonal phases of these materials, which have been seen in
epitaxially grown thin films, must be stabilized by the substrate
and cannot be free-standing films.
\end{abstract}



\maketitle

\section{Introduction}
\label{sec:intro}

The concept of the tetragonal distortion of a face-centered cubic
lattice into a body-centered cubic lattice, as well as the reverse
transformation, was introduced by Bain in 1924\cite{bain24} as a
model for the formation of Martensite in steels.  Although this
model is not exactly correct,\cite{wayman95:cryst} the concept of
the ``Bain path'' has been useful in investigating the behavior of
element and compound metallic systems.  This usefulness has
increased with the development of total energy electronic structure
methods, which can investigate regions of the Bain path not
accessible by experimental methods.  By following the Bain path from
fcc to bcc, Mehl and Boyer\cite{mehl91:_calcu} (in the following
referred to as MB) showed that the fcc elements aluminum and iridium
are elastically unstable in the bcc phase, i.e., the Bain path
reaches a local maximum at the position of the bcc lattice.  Wills,
{\em et al.},\cite{wills92:cijtrend} did first-principles
calculations of the Bain path of gold and platinum which showed that
these elements were also unstable in the bcc phase.  In addition
they used the reverse Bain path, from bcc to fcc, to show that
tantalum and tungsten were elastically unstable in the fcc phase.
Craievich, {\em et
al.},\cite{craievich94:stability,craievich97:unstable} looked at the
fcc elements nickel and rhodium, which were unstable in the bcc
state, and the bcc elements cobalt, chromium, molybdenum, niobium,
tantalum, and vanadium, which were elastically unstable in the fcc
state.  Jona and Marcus showed that both bcc copper\cite{jona01:cu}
and palladium\cite{jona02:pd} are also elastically unstable along
the Bain path.  Even metals which are only metastable in the fcc
phase, e.g., cobalt, manganese, osmium, rhenium, ruthenium,
titanium, and
hafnium,\cite{craievich94:stability,marcus97:fccti,aguayo02:hcpasfcc}
are elastically unstable to Bain path distortions in the bcc phase.

The above lists demonstrate that for elements to the right of the
alkaline earths on the periodic table, a stable or metastable fcc
(bcc) phase implies an elastically unstable bcc (fcc) phase.  This,
however, does not end the search for metastable phases along the
Bain path.  As noted by Craievich, {\em et
al.},\cite{craievich94:stability} the existence of a minimum and a
local maximum in the Bain path implies the existence of another
local minimum.  This minimum must be outside the classical Bain path
from fcc to bcc, though this does not preclude the existence of yet
another minimum between fcc and bcc.  MB\cite{mehl91:_calcu} found a
local minimum in the extended Bain path of aluminum by compressing
along the [001] direction until $c/a = 0.567$, where $c/a = 1$ is
the fcc phase and $c/a = 1/\sqrt2$ the bcc phase.  Iridium was found
to have a minimum at $c/a = 0.566$.  These minima are very close to
the special value $c/a = 1/\sqrt3 \approx 0.577$, where the
body-centered tetragonal lattice is 10-fold coordinated
(bct$_{10}$).  Similar minima are found for
copper,\cite{jona01:cu,mishin01:cu} gold,\cite{kirchhoff01:autbmd}
palladium,\cite{jona02:pd} and fcc
titanium.\cite{craievich94:stability,sliwko96:fcc_to_bcc} This
crystal structure has been observed in the $\alpha$ phase of
protactinium and in the $\beta$ phase of
mercury,\cite{donohue74:elements} and was given the {\em
Strukturbericht} designation A$_a$.\cite{asm_vol3}

For bcc crystals, the local minimum is at $c/a > \sqrt2$, where $c/a
= 1$ is the bcc lattice and $c/a = \sqrt2$ is the fcc lattice.
Sliwko, {\em et al.},\cite{sliwko96:fcc_to_bcc} found a local
minimum in vanadium at $c/a \approx 1.8$.  There is no increased
coordination at this site.  Experimentally this structure is
observed in indium,\cite{donohue74:elements} and has the {\em
Strukturbericht} designation A6.\cite{asm_vol3}

Given the increased coordination near the secondary minimum in the
Bain path for fcc crystals, one is tempted to conclude that this
state is metastable.  This turns out not to be the case.  As found
by Boyer,\cite{boyer89:magic} there is an orthorhombic ``magic
strain'' (properly, Physically Allowed Lattice-Invariant (PALI)
strain) which transforms one fcc lattice into another fcc lattice.
A similar orthorhombic magic strain transforms one bcc lattice into
another.  MB showed that the magic strain path for fcc aluminum
passes through the secondary minimum of the aluminum Bain path, and
that, when viewed in the space of the orthorhombic strains, this
point is actually a saddle point and hence unstable.  In terms of
the elastic constants, $C_{11} - C_{12}$ for the secondary minimum
is negative.  Jona and Marcus demonstrated this instability by
direct calculation of the elastic constants of
copper\cite{jona01:cu} and palladium\cite{jona02:pd} at the
secondary Bain path minimum.

The concept of magic strains was generalized by Van de
Waal\cite{waal90:com_magic} and applied to the study of metastable
structures in various systems, including
silicon\cite{kaxiras94:magic} and model two dimensional
lattices.\cite{boyer96:2dmd} In this paper we use the formalism to
show that the secondary minimum along the Bain path of an elemental
fcc or bcc solid is unstable with respect to an orthorhombic shear.
We note that this need not be the case for more complicated unit
cells -- the metastable BCT5 state of
silicon\cite{boyer91:_newlo,kaxiras94:magic} was found using the
concept of magic strains.  The magic strain formalism is a useful
tool for searching for metastable structures.\cite{stokes02:trans}
Indeed, this paper can be considered to be a search for metastable
states due to orthorhombic distortions of stable fcc and bcc
crystals.  In this case, however, we will show that there are no
such states.

In Section~\ref{sec:bain}, we show an extension of the Bain path for
both fcc and bcc metals.  In Section~\ref{sec:ortho} we show the
construction of the orthorhombic strains required for fcc
$\rightarrow$ fcc and bcc $\rightarrow$ bcc magic strains and
explore large regions of strain space using parametrized fits to
first-principles results via the NRL Tight-Binding
method.\cite{mehl96:_appli} These results show that, within the
tight-binding model of fcc and bcc metals the Bain path minima
discussed above are elastically unstable.  In Section~\ref{sec:cij}
we employ first-principles density functional theory methods to
directly compute the elastic constants for structures at the
secondary Bain path minimum.  These more accurate results show that
such structures are indeed elastically unstable.  We conclude in
Section~\ref{sec:sum} with a brief summary and discussion of our
results.

\section{Extending the Bain Path}
\label{sec:bain}

We can represent the Bain path of an elemental compound by the
standard body-centered tetragonal unit cell,

\begin{equation}
\left(
\begin{array}{c}
{\bf a}_1 \\
{\bf a}_2 \\
{\bf a}_3
\end{array}
\right)
=
\left(
\begin{array}{ccc}
-\frac12 a & \frac12 a & \frac12 c \\
\frac12 a & -\frac12 a & \frac12 c \\
\frac12 a & \frac12 a & -\frac12 c \\
\end{array}
\right)
\cdot
\left(
\begin{array}{c}
\hat{x} \\
\hat{y} \\
\hat{z}
\end{array}
\right) ~ .
\label{equ:bctprim}
\end{equation}
If $c = a$ in (\ref{equ:bctprim}) we have a bcc lattice, while if $c
= \sqrt2 a$ we have an fcc lattice.  Alternatively, we can represent
the Bain path using a {\em face}-centered cubic unit cell,
\begin{equation}
\left(
\begin{array}{c}
{\bf a}_1 \\
{\bf a}_2 \\
{\bf a}_3
\end{array}
\right)
=
\left(
\begin{array}{ccc}
0 & \frac12 a' & \frac12 c \\
\frac12 a' & 0 & \frac12 c \\
\frac12 a' & \frac12 a' & 0 \\
\end{array}
\right)
\cdot
\left(
\begin{array}{c}
\hat{x} \\
\hat{y} \\
\hat{z}
\end{array}
\right) ~ ,
\label{equ:fctprim}
\end{equation}
where
\begin{equation}
a' = \sqrt2 a ~ .
\label{equ:acvrt}
\end{equation}
Thus when $c = a'$ we have an fcc lattice, while when $c =
a'/\sqrt2$ we have a bcc lattice.  We stress that
(\ref{equ:bctprim}) and (\ref{equ:fctprim}) are different
representations of the same lattice.  In this case, the unit cell
represented by (\ref{equ:fctprim}) is related to the one represented
by (\ref{equ:bctprim}) by a rotation of 45$^\circ$ about the $z$
axis.

\begin{figure}
\includegraphics[width=3.4in]{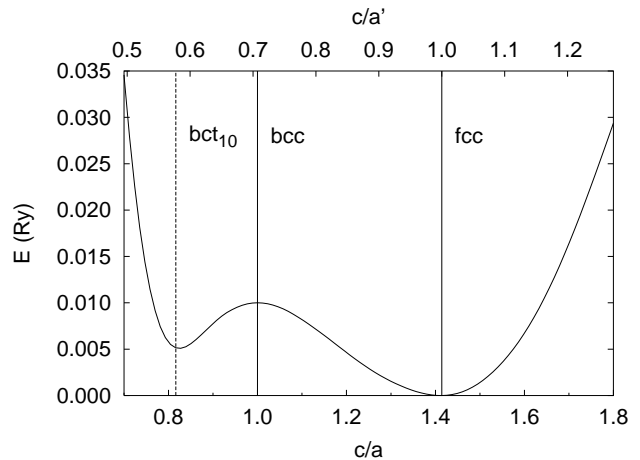}
\caption{\label{fig:ptbain}Bain path for platinum as a function of
$c/a$ in (\protect\ref{equ:bctprim}), using the tight-binding
parameters described in the text.  The volume of the unit cell was
chosen to minimize the total energy at each value of $c/a$.  The
lines indicate the positions of the bct$_{10}$, bcc, and fcc states.
The top $x$-axis shows the corresponding value of $c/a'$ from
(\protect\ref{equ:fctprim}).}
\end{figure}

To illustrate our techniques we begin with platinum, which has an
equilibrium fcc structure.  Fig.~\ref{fig:ptbain} shows the energy
along the platinum Bain path as a function of $c/a$.  The total
energy of the system was evaluated using the NRL tight-binding
parameters for platinum\cite{mehl96:_appli,notes:ptget}.  The
Brillouin zone was sampled by using a regular mesh of $k$-points
generated at $c/a = 1$, i.e., the bcc lattice, but using the
body-centered tetragonal symmetry of the system (space group
$I4/mmm-D_{4h}^{17}$).  This turns out to be the most efficient way
to generate $k$-points for all of the strains covered in this
calculation.  It does, however, require a large number of mesh
points.  We used 2119 points in the irreducible part of the
Brillouin zone, and carefully checked our results using other
k-point meshes to insure convergence.\cite{mehl00:_occup} The total
energy of the system was computed using the method described by
Gillan\cite{gillan89:alvac}, with a Fermi broadening temperature of
5~mRy.  By comparing with smaller k-point meshes, we estimate that
the numerical error in the calculation of the total energy was less
than 0.1~mRy for $c/a < 1.5$, and less than 0.2~mRy over the entire
range of Fig.~\ref{fig:ptbain}.

As expected, the global minimum in Fig.~\ref{fig:ptbain} is at the
fcc position, $c/a = \sqrt2$.  (Obtaining this result was considered
a necessary condition to show k-point convergence.)  As for other
fcc elements,\cite{mehl91:_calcu,jona01:cu,jona02:pd} there is a
local minimum near the bct$_{10}$ structure, which is at $c/a =
\sqrt{2/3} \approx 0.816$ using representation (\ref{equ:bctprim}).
Both protactinium and the $\beta$ phase of mercury exist in stable
or metastable configurations near this bct$_{10}$
structure.\cite{donohue74:elements}

\begin{figure}
\includegraphics[width=3.4in]{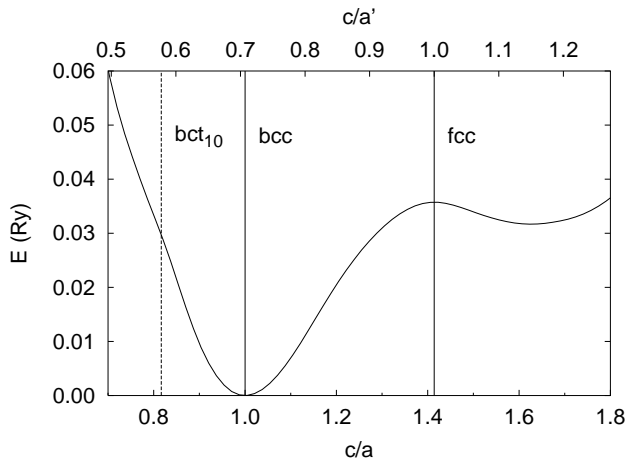}
\caption{\label{fig:wbain}Bain path for tungsten using the
tight-binding parameters and $k$-points described in the text.  The
notation is the same as in Fig.~\protect\ref{fig:ptbain}.}
\end{figure}

In a similar way we can study the extension of the Bain path for an
element with a bcc ground state.  This is most easily done starting
with the face-centered primitive vectors (\ref{equ:fctprim}),
constructing a regular k-point mesh at $c = a'$, (the fcc lattice),
again using space group $I4/mmm-D_{4h}^{17}$.  We use a mesh with
4069 points in the irreducible part of the Brillouin zone.
Figure~\ref{fig:wbain} shows our results for tungsten, using our
tight-binding parameters.\cite{mehl96:_appli,notes:wget} We note
that there is a local minimum near $c/a' = 1.15$.  This minimum is
required by the continuity of $E(c/a)$, since we know that the fcc
structure of tungsten is elastically unstable and so a local
maximum.  Unlike the case of the strained fcc elements, there is no
high-coordination structure near this point.  Interestingly, we do
see a small shoulder in the energy-strain curve at the location of
the bct$_{10}$ structure.  Experimentally, Indium exists in this
structure, with $c/a' = 1.08$.\cite{donohue74:elements}

\section{Orthorhombic Strains}
\label{sec:ortho}

In the previous section we saw that the extension of the Bain path
has a secondary minimum for elements with either fcc or bcc ground
states.  It is natural to ask whether or not this secondary minimum
is in fact metastable.  Here we show a procedure, based on
MB,\cite{mehl91:_calcu} which answers this question, and, in the
process shows that there is no evidence such a metastable state
exists for any element at zero temperature.

\begin{figure}
\includegraphics[width=3.4in]{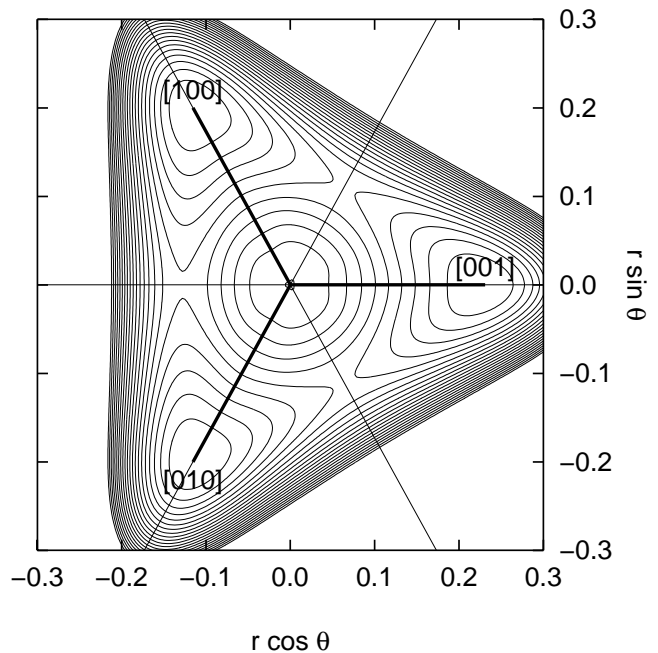}
\caption{\label{fig:ptmap}Map of the total energy of the lattice
described by (\protect\ref{equ:bcoprim}) for platinum, using the NRL
tight-binding parameters, and choosing the minimum energy volume at
each point.  The heavy straight lines represent the Bain path from
bcc (at the origin) to one of three fcc lattices.  The lighter
straight lines represent the extended Bain path.  The labels
indicate the direction of the tetragonal lattice along the Bain
paths.  The fcc lattice points are the minima near the labels.  The
bcc lattice point, at the origin, is 10~mRy above the energy of the
fcc lattice, and the contour interval is 1~mRy.  The coordinates
$(r,\theta)$ are described in (\protect\ref{equ:bcoxyz}).}
\end{figure}

\subsection{Elements with an fcc ground state}
\label{subsec:fccg}

The first check in determining the metastability of the bct
structure in Fig.~\ref{fig:ptbain} would be to look at
long-wavelength distortions of the unit cell, i.e., the elastic
constants.  We will, in fact, use the technique outlined in MB which
will allow us to look at general strains about the bct position.

We begin by writing the primitive vectors for an arbitrary
body-centered orthorhombic unit cell in a form reminiscent of
(\ref{equ:bctprim}):

\begin{equation}
\left(
\begin{array}{c}
{\bf a}_1 \\
{\bf a}_2 \\
{\bf a}_3
\end{array}
\right)
=
\left(\frac{V}{4}\right)^\frac13
\left(
\begin{array}{ccc}
-x &  y &  z \\
 x & -y &  z \\
 x &  y & -z \\
\end{array}
\right)
\cdot
\left(
\begin{array}{c}
\hat{x} \\
\hat{y} \\
\hat{z}
\end{array}
\right) ~ ,
\label{equ:bcoprim}
\end{equation}
where $V$ is the volume of the unit cell, and
\begin{eqnarray}
x & = & \exp [r \cos(\theta + 2 \pi/3)] \nonumber \\
y & = & \exp [r \cos(\theta - 2 \pi/3)] ~ , ~ \mbox{and} \nonumber \\
z & = & \exp [r \cos \theta] ~ ,
\label{equ:bcoxyz}
\end{eqnarray}
so that $xyz = 1$ for all values of $(r,\theta)$.  We immediately
see that the energy of this unit cell, $E(V,r,\theta)$, is periodic
in $\theta$ with period $2\pi/3$, and that $r = 0$ represents the
bcc lattice for all values of $\theta$.  Furthermore, in the special
cases $\theta = (0, 2\pi/3, -2\pi/3)$ the orthorhombic lattice
becomes tetragonal, with the tetragonal axis in the $z$, $x$, or $y$
direction, respectively.  These lines are just the extended Bain
path of the system, stretched along the [001], [100], or [010]
direction.  We find the face-centered cubic lattice when $r = (\ln
2)/3$ along any of these paths.

We used our platinum tight-binding parameters to map the energy
surface $E(V,r,\theta)$ of (\ref{equ:bcoprim}), minimizing the total
energy with respect to volume at each point $(r,\theta)$.  The space
group is now {\em Immm}$-D_{2h}^{25} (\#71)$, and we used a mesh of
3925 $k$-points in the irreducible Brillouin zone.  The results are
shown in Fig.~\ref{fig:ptmap}.  In particular, we note that along
each Bain path, there is a saddle point about 5~mRy above the fcc
ground state.  These points represent the same lattice as the local
minimum structure in Fig.~\ref{fig:ptbain}.  We see that this point
is unstable with respect to an orthogonal distortion, which means
that a least one linear combination of the lattice constants of the
tetragonal structure must be negative.  To see which one, let $a_0$
and $c_0$ be the lattice constants representing the secondary
minimum in the Bain path using the primitive vectors
(\ref{equ:bctprim}).  For a crystal with an fcc ground state and a
maximum at the bcc position, we will have $c_0/a_0 < 1$, though this
is not necessary for the following analysis.  If we use
(\ref{equ:bcoprim}) and (\ref{equ:bcoxyz}) to represent this
structure we have
\begin{eqnarray}
V & = V_0 & = 1/2 \, a_0^2 c_0 ~ , \nonumber \\
r & = r_0 & = 2/3 \, \ln c_0/a_0 ~ , ~ \mbox{and} \nonumber \\
\theta & = \theta_0 & = 0 ~ .
\label{equ:bcoequ}
\end{eqnarray}
Let us fix the volume and expand $r = r_0 + 2 \alpha$ and $\theta$
about the equilibrium point.  Then to first order in $\alpha$ and
$\theta$, (\ref{equ:bcoprim}) becomes
\begin{widetext}
\begin{equation}
\left(
\begin{array}{c}
{\bf a}_1 \\
{\bf a}_2 \\
{\bf a}_3
\end{array}
\right)
=
\left(
\begin{array}{rrr}
- \frac12 a_0 ( 1 - \alpha + \beta) &
 \frac12 a_0 ( 1 - \alpha - \beta) &
 c_0 ( 1 + 2 \alpha) \\
 \frac12 a_0 ( 1 - \alpha + \beta) &
- \frac12 a_0 ( 1 - \alpha - \beta) &
 c_0 ( 1 + 2 \alpha) \\
 \frac12 a_0 ( 1 - \alpha + \beta) &
 \frac12 a_0 ( 1 - \alpha - \beta) &
-  c_0 ( 1 + 2 \alpha) \\
\end{array}
\right)
\cdot
\left(
\begin{array}{c}
\hat{x} \\
\hat{y} \\
\hat{z}
\end{array}
\right) ~ ,
\label{equ:bcoplin}
\end{equation}
\end{widetext}
where
\begin{equation}
\beta = \sqrt3 r_0 \theta/2 ~ .
\label{equ:qdef}
\end{equation}
In terms of the standard elastic strain parameters for the
body-centered tetragonal lattice we have
\begin{eqnarray}
e_1 = -\alpha + \beta ~ , &
e_2 = -\alpha - \beta ~ , &
e_3 = 2 \alpha ~ , \nonumber \\
& e_4 = e_5 = e_6 = 0 & ~ .
\label{equ:bcostrain}
\end{eqnarray}
The lowest order contribution to the elastic energy associated with
this strain is
\begin{equation}
\Delta E = V [(C_{11} + C_{12} - 4 C_{13} + 2 C_{33})
\alpha^2 + (C_{11} - C_{12}) \beta^2] ~ .
\label{equ:bcoelastic}
\end{equation}
If this structure was stable, the coefficients of the $\alpha^2$ and
$\beta^2$ terms should be positive, in agreement with Born and
Huang's derivation of the elastic stability
criteria.\cite{born54:dynamics} We note that keeping $\theta = 0$
and varying $\alpha$ moves the lattice along the Bain path.  Since
the point $\alpha = 0$ is a minimum along that path, the first
linear combination of elastic constants must be positive.  However,
as seen in Fig.~\ref{fig:ptmap}, fixing $\alpha = 0$ and varying
$\theta$ (and therefore $\beta$) lowers the energy, implying that
$C_{11} - C_{12} < 0$.

\subsection{Elements with a bcc ground state}
\label{subsec:bccg}

Our examination of the stability of the bct structure
(Fig.~\ref{fig:wbain}) of an element with a bcc ground state is
similar to the above discussion, with minor differences because we
are now describing a state with $c/a$ expanded from the equilibrium,
rather than compressed as it is for fcc elements.

We therefore begin with an orthorhombic distortion of the fcc-like
primitive vectors (\ref{equ:fctprim}):

\begin{equation}
\left(
\begin{array}{c}
{\bf a}_1 \\
{\bf a}_2 \\
{\bf a}_3
\end{array}
\right)
=
\left(\frac{V}{2}\right)^\frac13
\left(
\begin{array}{ccc}
 0 & y & z \\
 x & 0 & z \\
 x & y & 0 \\
\end{array}
\right)
\cdot
\left(
\begin{array}{c}
\hat{x} \\
\hat{y} \\
\hat{z}
\end{array}
\right) ~ ,
\label{equ:fcoprim}
\end{equation}
where $V$ is the volume of the unit cell and
\begin{eqnarray}
x & = & \exp [- r \cos(\theta + 2 \pi/3)] \nonumber \\
y & = & \exp [- r \cos(\theta - 2 \pi/3)] ~ , ~ \mbox{and} \nonumber \\
z & = & \exp [- r \cos \theta] ~ .
\label{equ:fcoxyz}
\end{eqnarray}
In analogy with (\ref{equ:bcoprim}), there are three Bain paths
along the directions $\theta = (0, \pm 2\pi/3)$, but now $r = 0$ is
an fcc lattice, and at the special value $r = (\ln 2)/3$ along any
one of the Bain paths we find a bcc lattice.

\begin{figure}
\includegraphics[width=3.4in]{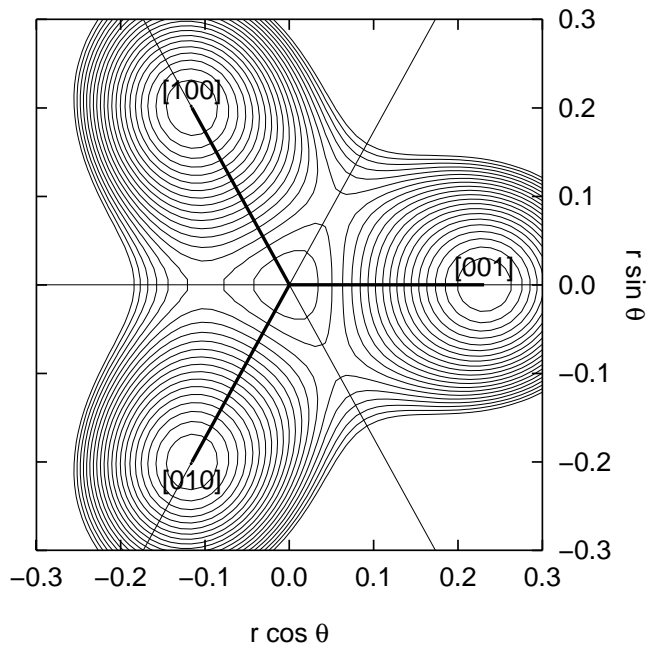}
\caption{\label{fig:wmap}
Map of the total energy of the lattice (\protect\ref{equ:fcoprim})
for tungsten, using the NRL tight-binding parameters, choosing the
minimum energy volume at each point.  The origin of the graph is the
fcc lattice, the heavy straight lines radiating from the lattice are
the Bain paths, and the lighter straight lines are the extensions of
the Bain path.  The labels indicate the orientation of the
tetragonal axis along each Bain path.  The contour interval is
2~mRy, and the energy of the fcc structure is 35.8~mRy above the
energy of the bcc structure.  The coordinates $(r,\theta)$ are
described in (\protect\ref{equ:fcoxyz}).}
\end{figure}

Using the tungsten tight-binding parameters, we choose to describe
the space group for (\ref{equ:fcoxyz}) as {\em Fmmm}$-D_{2h}^{23}
(\#69)$.  We use a mesh of 7813 $k$-points in the irreducible part
of the Brillouin zone.  The results of the calculation are shown in
Fig.~\ref{fig:wmap}.  The saddle points located along the extended
Bain path, approximately 30~mRy above the ground state, represent
the same structure as the local minimum found in
Fig.~\ref{fig:wbain}.  We can determine the nature of the elastic
instability here in a manner similar to Sec.~\ref{subsec:fccg}.  Let
$a_0'$ and $c_0$ specify the local minimum on the Bain path in
Figs.~\ref{fig:wbain} and \ref{fig:wmap}, using the primitive
vectors (\ref{equ:fctprim}).  In terms of (\ref{equ:fcoprim}), this
occurs at
\begin{eqnarray}
V & = V_0 & = 1/4 \, a_0'^2 c_0 ~ , \nonumber \\
r & = r_0 & = - 2/3 \, \ln c_0/a_0' ~ , ~ \mbox{and} \nonumber \\
\theta & = \theta_0 & = 0 ~ .
\label{equ:fcoequ}
\end{eqnarray}
Taking $r = r_0 + 2 \alpha$ and expanding the lattice parameters in
powers of $\alpha$ and $\theta$, we find that (\ref{equ:fcoprim})
can be written as an orthorhombic distortion of (\ref{equ:fctprim}).
To lowest order, the elastic strain parameters are
\begin{eqnarray}
e_1 = \alpha + \beta ~ , &
e_2 = \alpha - \beta ~ , &
e_3 = - 2 \alpha ~ , \nonumber \\
& e_4 = e_5 = e_6 = 0 & ~ ,
\label{equ:ccostrain}
\end{eqnarray}
where $\beta$ is defined in (\ref{equ:qdef}).  Thus the elastic energy
associated with this strain is identical to (\ref{equ:bcoelastic}),
and, once again, we see that at the Bain path minimum $C_{11} -
C_{12} < 0$.

Traditionally, elastic constants in centered-tetragonal lattices are
computed using the body-centered form of the primitive vectors
(\ref{equ:bctprim}) rather than the face-centered form
(\ref{equ:fctprim}).  As we noted, this corresponds to a 45$^\circ$
rotation of the coordinates around the $z$ axis.  Under this
rotation the elastic constants transform so that\cite{mehl90:_struc}
\begin{equation}
\left(
\begin{array}{c}
C_{11} + C_{12} \\
C_{33} \\
C_{13} \\
C_{44} \\
C_{11} - C_{12} \\
2 C_{66}
\end{array}
\right)_{\mbox{fct}}
\rightarrow
\left(
\begin{array}{c}
C_{11} + C_{12} \\
C_{33} \\
C_{13} \\
C_{44} \\
2 C_{66} \\
C_{11} - C_{12}
\end{array}
\right)_{\mbox{bct}} ~ .
\label{equ:cijtrans}
\end{equation}
When we compute energy change of the strain (\ref{equ:ccostrain}) on
the structure defined by (\ref{equ:fcoequ}) using the body-centered
tetragonal primitive vectors (\ref{equ:bctprim}), then, the energy
changes by
\begin{equation}
\Delta E = V [(C_{11} + C_{12} - 4 C_{13} + 2 C_{33})
\alpha^2 + 2 C_{66} \beta^2] ~ .
\label{equ:fcoelastic}
\end{equation}
The saddle point behavior of the energy surface near this minimum
leads to the prediction that $C_{66} < 0$.

\begin{figure}
\includegraphics[width=3.2in]{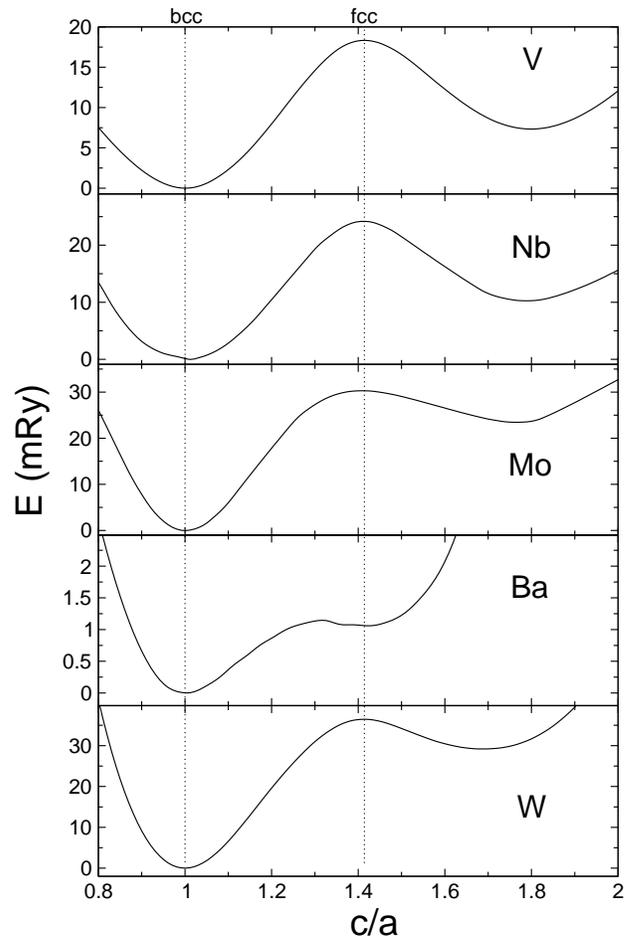}
\caption{\label{fig:bccbain}Total energy as a function of $c/a$
at the equilibrium volume for elements with a bcc ground state.}
\end{figure}

\section{Elastic constants of the structures at the Bain local
minima}
\label{sec:cij}

In the previous section we showed that for typical fcc and bcc
structured elements the secondary minima in the Bain path, as shown
in Figs.~\ref{fig:ptbain} and \ref{fig:wbain}, are actually saddle
points in the space defined by the orthorhombic unit cells
(\ref{equ:bcoprim}) and (\ref{equ:fcoprim}).  From
Figs.~\ref{fig:ptmap} and \ref{fig:wmap} we see that the only
elastically stable structures along the Bain path are the fcc and
bcc lattices.  In rare cases both of these structures may be minima
on the Bain path.  In this section we will check this conjecture via
first principles density functional theory calculations using the
Linearized Augmented Plane Wave (LAPW)
method\cite{andersen75:_linea} as implemented in the Wien
code\cite{blaha02:wien2k} with the generalized gradient
approximation (GGA) exchange correlation functional of Perdew,
Burke, and Ernzerhof.\cite{perdew96:pbegga}

\begin{table}
\caption{\label{tab:bainmin}Position of the secondary minimum along
the extended Bain path, $c/a$, for the elemental metals which have
fcc or bcc ground states.  We also show the energy difference
between the phases.  All calculations were done by minimizing the
total energy as a function of volume for fixed $c/a$, using the
primitive vectors in (\protect\ref{equ:bctprim}).  Energy
differences are in mRy.  Note that for strontium the secondary
minimum coincides with the bcc lattice, and for barium the minimum
coincides with the fcc lattice.}
\begin{ruledtabular}
\begin{tabular}{cc....}
Metal & Equil. & \multicolumn{1}{r}{$c/a$} &
 \multicolumn{1}{r}{$|E_{\rm bcc}-E_{\rm fcc}|$} &
 \multicolumn{1}{r}{$|E_{\rm bcc}-E_{\rm bct}|$} &
 \multicolumn{1}{r}{$|E_{\rm fcc}-E_{\rm bct}|$} \\
\hline
Al & fcc & 0.814 & 7.287  & 1.544  & 5.743  \\
Ag & fcc & 0.894 & 2.404  & 0.164  & 2.240  \\
Au & fcc & 0.852 & 1.785  & 0.307  & 1.478  \\
Ba & bcc & 1.414 & 1.060  & 1.060  & 0.000  \\
Ca & fcc & 0.911 & 1.648  & 0.175  & 1.473  \\
Cu & fcc & 0.935 & 2.706  & 0.033  & 2.673  \\
Ir & fcc & 0.801 & 48.997 & 23.684 & 25.313 \\
Mo & bcc & 1.765 & 30.301 & 23.450 & 5.851  \\
Nb & bcc & 1.787 & 24.186 & 10.245 & 13.941 \\
Pd & fcc & 0.876 & 3.752  & 0.405  & 3.347  \\
Rh & fcc & 0.814 & 26.468 & 10.644 & 15.824 \\
Sr & fcc & 0.707 & 0.730  & 0.000  & 0.730  \\
V  & bcc & 1.799 & 18.330 & 7.332  & 10.998 \\
W  & bcc & 1.687 & 36.471 & 29.195 & 7.276  \\
\end{tabular}
\end{ruledtabular}
\end{table}

\begin{table}
\caption{\label{tab:cij}Elastic constants of at the secondary
minimum in the Bain path for the elements described in
Table~\ref{tab:bainmin}, with $C' = (C_{11}-C_{12})/2$.  All elastic
constants are computed using the orientation and unit cell described
by (\protect\ref{equ:bctprim}), and are expressed in GPa.}
\begin{ruledtabular}
\begin{tabular}{ccrr}
Metal & Equil. & $C'$ & $C_{66}$ \\
\hline
Al & fcc & -18 & 127 \\
Ag & fcc & -56 & 133 \\
Au & fcc & -40 & 177 \\
Cu & fcc & -11 & 227 \\
Ir & fcc & -69 & 689 \\
Mo & bcc & 94 & -112 \\
Nb & bcc & 82 & -65 \\
Pd & fcc & -6 & 241 \\
Rh & fcc & -36 & 498 \\
V & bcc & 53 & -39 \\
W & bcc & 75 & -140 \\
\end{tabular}
\end{ruledtabular}
\end{table}

We begin by determining the position of the secondary minimum in the
Bain path for cubic elemental metals.  In Fig.~\ref{fig:bccbain} we
show the results for several of the elemental bcc metals.  In all
cases there are two minima in the calculation, one at the bcc site
and one at or beyond the fcc position on the Bain path.  The results
are shown in Table~\ref{tab:bainmin}, where the ratio $c/a$ is
determined using the representation of the body-centered tetragonal
lattice in (\ref{equ:bctprim}).  With the exceptions of barium and
strontium, all of the metals shown here have a secondary minimum of
the type described in Sec.~\ref{sec:bain}:  if the ground state is
fcc, the minimum is at $c/a < 1$, while if the ground state is bcc,
the minimum is at $c/a > \sqrt2$.  Barium and strontium are unique
in that the secondary minimum is at the fcc point on the Bain path
for barium, as shown in Fig.~\ref{fig:bccbain}, and the at the bcc
point for strontium.  We will not consider these elements further in
this paper.

For the remaining elements we compute the elastic constants of the
tetragonal system using standard
techniques,\cite{mehl90:_struc,jona01:cu} using the primitive
vectors (\ref{equ:bctprim}) to describe the lattice.  We present the
results of our calculations in Table~\ref{tab:cij}.  As predicted,
all of the fcc ground state elements have $C' = (C_{11} - C_{12})/2
< 0$ at the secondary minimum, and all bcc ground state elements
have $C_{66} < 0$, in agreement with the predictions of
Sec.~\ref{sec:ortho}.

\section{Summary}
\label{sec:sum}

As noted in the introduction, it has previously been shown that the
secondary minimum is elastically unstable in
aluminum,\cite{mehl91:_calcu} copper,\cite{jona01:cu}
iridium,\cite{mehl91:_calcu} and palladium.\cite{jona02:pd} To our
best knowledge the stability or instability of the secondary minimum
of a bcc element has never before been determined.  The calculations
done here show that if the energy along the Bain path for a metal is
at a minimum at the fcc (bcc) structure and a maximum at the bcc
(fcc) structure, then there is a secondary minimum with a bct
structure with is elastically unstable.

The implication of this work is that there can be no free standing
body-centered tetragonal structures for these metals.  Any bct
structures observed, such as in thin films\cite{alippi197:bainpath}
must be stabilized by an external force.  In the case of epitaxially
grown thin films, this is force is presumably supplied by the
substrate.  For sufficiently thick films this supporting force will
be insufficient to stabilize the bct structure, and the film will
revert to its ground state, as Bencok {\em et
al.}\cite{bencok98:vonfe} observed for thin vanadium films grown on
Fe(100).

While we have not discussed the case of hcp metals such as titanium,
where epitaxial bct structures are
seen,\cite{marcus97:fccti,kim96:tional} we note that the minimum
energy bcc structure of Ti is known to be elastically
unstable.\cite{sliwko96:fcc_to_bcc} The conclusions of this paper
thus apply to Ti as well.

\begin{acknowledgments}
This work was supported by the U. S. Office of Naval Research.  The
development of the tight-binding codes was supported in part by the
U. S. Department of Defense Common HPC Software Support Initiative
(CHSSI).  One of us (RdC) is partially supported by CONACYT-M\'exico
under Grant Nos. 34501-E and 43830-F.
\end{acknowledgments}

\end{document}